\documentclass[12pt]{article}
\usepackage[utf8]{inputenc}
\usepackage{amsmath}
\usepackage{natbib}
\usepackage{xcolor}
\usepackage{graphicx}
\usepackage{appendix}
\usepackage{tablefootnote}
\usepackage{longtable}
\usepackage{adjustbox}
\usepackage{pifont} 
\usepackage{hyperref}

\newcommand{\blind}{1}

\begin{document}

\def\spacingset#1{\renewcommand{\baselinestretch}%
{#1}\small\normalsize} \spacingset{1}

\if1\blind
{
  \title{\bf Thirty Years of The Network Scale-up Method\footnote{This article has been accepted for publication in the Journal of the American Statistical Association, published by Taylor \& Francis. This is the original version of the manuscript. The updated manuscript can be found at \url{https://doi.org/10.1080/01621459.2021.1935267}}}
  \author{Ian Laga, Le Bao, and Xiaoyue Niu \\
    Department of Statistics, Pennsylvania State University}
    \date{}
  \maketitle
} \fi

\if0\blind
{
  \bigskip
  \bigskip
  \bigskip
  \begin{center}
    {\LARGE\bf Thirty Years of The Network Scale-up Method}
\end{center}
  \medskip
} \fi

\bigskip
\begin{abstract}
   Estimating the size of hard-to-reach populations is an important problem for many fields. The Network Scale-up Method (NSUM) is a relatively new approach to estimate the size of these hard-to-reach populations by asking respondents the question, ``How many X's do you know,'' where X is the population of interest (e.g. ``How many female sex workers do you know?''). The answers to these questions form Aggregated Relational Data (ARD). The NSUM has been used to estimate the size of a variety of subpopulations, including female sex workers, drug users, and even children who have been hospitalized for choking. Within the Network Scale-up methodology, there are a multitude of estimators for the size of the hidden population, including direct estimators, maximum likelihood estimators, and Bayesian estimators. In this article, we first provide an in-depth analysis of ARD properties and the techniques to collect the data. Then, we comprehensively review different estimation methods in terms of the assumptions behind each model, the relationships between the estimators, and the practical considerations of implementing the methods. Finally, we provide a summary of the dominant methods and an extensive list of the applications, and discuss the open problems and potential research directions in this area. 
\end{abstract}

\noindent%
{\it Keywords:} Size estimation, small area estimation, key populations, aggregated relational data.
\vfill

\newpage
\spacingset{1.5} 
\section{Introduction}
\label{sec:intro} 
Estimating the size of hard-to-reach populations is an important problem in a variety of contexts. Governments and humanitarian organizations which aim to eradicate infectious diseases and improve the lives of citizens through treatment programs are interested in population sizes because the treatment target needs to be clear and funds need to be allocated correctly. The Joint United Nations Programme on HIV and AIDS (UNAIDS) aims to limit the spread of HIV by locating large HIV populations. Groups that are particularly vulnerable are known as target- or key-populations. For example, female sex workers (FSW) are among the highest subpopulation living with HIV. It is difficult to estimate the size of FSW directly because of social stigma around sex work and because FSW comprise a relatively small percentage of the general population. Existing approaches to estimate these hard-to-reach populations include mark-recapture, mapping, and venue-based surveys. See \cite{bernard2010counting} for a detailed list of population size estimation methods and \cite{sabin2016availability} for a comparison of the availability and quality of different data types when estimating certain key subpopulations.

A relatively new method for estimating the size of key populations is the Network Scale-up Method (NSUM), based on the basic scale-up model \citep{johnsen1989estimating}. The authors were in Mexico soon after an earthquake and were interested in estimating the number of people who had died in the earthquake. In this case, where the target population is people who have died from an earthquake, many existing methods were impossible to implement. One author asked people around Mexico City how many people they knew who had died in the earthquake. By leveraging only the responses about how many people each respondent knows, they were able to estimate the number of people who died from the earthquake. The NSUM uses questions from ``How many X's do you know?'' surveys to estimate both average network size and subpopulation sizes.

The method provides a relatively cheap, easy, and powerful tool for researchers and can still be applied when it is impossible or difficult to reach the target population directly. Furthermore, the NSUM respects the privacy of the respondents since respondents are not asked about their own characteristics or identifying information of their alters. UNAIDS and WHO published guidelines for the NSUM and noted that the NSUM is advantageous because respondents do not need to reveal their own status, questions can be added to existing household surveys, and sizes of multiple subpopulations can be estimated from one survey \citep{unaids2010guidelines}. Furthermore, obtaining data through NSUM can be $70-80\%$ cheaper than collecting traditional network data \citep{breza2017using}. However, due to biases from violating the assumptions of method, the UNAIDS and WHO designated the NSUM as a method ``under-development'' for estimating hard-to-reach populations \citep{unaids2010guidelines}, prompting the need for further development.

Here we standardize the notation we will use for the rest of the paper and provide a brief introduction to relevant network and NSUM terminology and notation to streamline reading of later sections. Aggregated relational data (ARD) refers to the data collected using ``How many X's do you know?'' questions, while NSUM refers to the process of estimating network size or subpopulation size using ARD. Let $N$ be the size of the general population, $N_k$ be the size of subpopulation $k$, and $n$ be the number of respondents. We denote the ARD responses by $y_{ik}$, the number of people that respondent $i$ reports knowing in subpopulation $k$. The network size, or degree, for person $i$ is $d_i$. Let $L$ be the number of subpopulations with known sizes. For simplicity, we assume only one subpopulation size is unknown, denoted by $N_u$, but stress that any number of subpopulations can be unknown in practice. Thus, the responses from person $i$ are denoted $y_{i\cdot} = (y_{i1}, \ldots, y_{iL}, y_{iu})$. The respondents to ARD surveys are also known as the ego and the people to whom the ego can form ties are called alters \citep{salganik2011game}.

The basic idea behind NSUM relies on the assumption that the proportion of the subpopulation to the general population is equal to the proportion of the person's network that belongs to the subpopulation, i.e.
\begin{equation}
\label{eq:basic}
    \frac{y_{ik}}{d_i} = \frac{N_k}{N}
\end{equation}
If the degree $d_i$ was known, then it would be easy to solve for $N_k$ using Equation (\ref{eq:basic}). However, estimating the degree is a difficult problem in and of itself \citep{rogerson1997estimating, dasgupta2014estimating}. A related topic is that of the ``small-world problem,'' which states that only a small number of connections can connect any two people \citep{rogerson1997estimating}. Estimating the average network size is also difficult because it is nearly impossible for someone to recount their entire social network without substantial effort and network size varies dramatically between individuals.

The model in Equation (\ref{eq:basic}) works well under three strong conditions: 1. everyone in the population is equally likely to know someone in subpopulation $k$, 2. for every person in a respondent's network, the respondent knows every subpopulation the person belongs to, and 3. respondents are able to fully recall everyone in their social network in the allotted time. These assumptions are commonly violated in practice. Barrier effects cause individuals to be more or less likely to know individuals in certain subpopulations, violating condition 1. Transmission errors block individuals from knowing everything about people in their social networks, violating condition 2. Recall errors result when people are unable to quickly count or remember everyone in their social network that belongs to a certain group, violating condition 3 \citep{johnsen1989estimating, johnsen1995social, killworth1998estimation, mccarty2001comparing}. We further explore each of these violations in later sections.


There are currently two software packages to analyze ARD using the NSUM, both implemented in \textsf{R} \citep{rlang}. The first, \texttt{NSUM} \citep{maltiel2015NSUM}, implements the models proposed by \cite{maltiel2015estimating}. The second, \texttt{networkreporting}  \citep{feehan2014networkreporting}, fits the generalized network scale-up model proposed by \cite{feehan2016generalizing}.

The rest of this manuscript is organized as follows. First, we offer a background to ARD and explore the features of the data and problems that arise when collecting ARD. Then, in Section \ref{sec:models}, we provide introductions to all significant NSU models. This section is divided into three subsections. In Section \ref{sec:basic}, we discuss the frequentist network scale-up estimators. These basic estimators are the most frequently used in practice due to their ease of use, but also include more complex models with increased flexibility. In Section \ref{sec:bayesian}, we introduce the Bayesian estimators. These models are reported to have improved the basic methods by better accounting for the sources of bias, but are also more difficult to use since they rely on Bayesian sampling algorithms. In Section \ref{sec:complete_net}, we review recent estimators that estimate complete network properties using only ARD. After discussing specific models, Section \ref{sec:adjustments} introduces modifications to model estimates. These approaches recognize limitations of the modeling procedures and further calibrate estimates using empirical studies. Final discussion is found in Section \ref{sec:discussion}.

\section{Properties of Aggregated Relational Data}
\label{sec:considerations}
Before discussing the various NSUM models, it is important to look at the properties of ARD in more detail. Specifically, there are four key biases that plague ARD: 1) Transmission errors, 2) Barrier effects, 3) Recall errors, and 4) Response biases. As with any survey, ARD can suffer from poor sampling behavior. Recall errors and response biases depend on the survey implementation, while transmission errors and barrier effects depend on the subpopulations and respondents.

\subsection{Transmission Error}
A response suffers from transmission error when the respondent is unaware that someone in their network belongs to a subpopulation \citep{killworth2006investigating}. The transmission error violates the assumption that respondents have perfect knowledge of which subpopulations their alters belong to and varies widely between different subpopulations \citep{killworth2006investigating, zheng2006many, maltiel2015estimating}.

\cite{shelley1995knows} studied transmission error by interviewing respondents in key subpopulations.  
HIV-positive respondents reported that only 49\% of their relatives were aware of their HIV-status. However, note that transmission error also exists for easy-to-reach populations. A large percentage of diabetics, twins, Native Americans, and widows and widowers all reported not revealing their status to some members of their social network \citep{killworth2006investigating}. Focus has primarily been on accounting for the transmission rate of the hard-to-reach subpopulations, although ignoring the transmission rate of the known subpopulation may also significantly influence estimates.

In some cases, researchers can estimate the \textit{transmission rate}, $\tau$, of a subpopulation. Also frequently called the \textit{visibility factor}, $\tau$ represents the fraction of a respondent's network that is aware the respondent is in the hidden subpopulation. The methods to estimate the visibility factor include expert opinion, comparison of NSU with proxy respondent method, social respect, coming-out ratio, and game of contacts \citep{haghdoost2018review}. Thus, in addition to collecting the ARD, researchers can also collect additional data to estimate the visibility factor. To the best of our knowledge, there has been no study that compares the accuracy of the different methods to estimate the visibility factor.

\subsection{Barrier Effect}
Barrier effects violate the constant proportion assumption because respondents can be more or less likely to know someone in a subpopulation due to their own characteristics, and are not limited to hard-to-reach populations. \citep{shelley1995knows, killworth2006investigating, shelley2006knows, salganik2011assessing}. For example, respondents were more likely to know people with whom they shared their race. 

The reasons for barrier effects can be both geographical as well as social. For example, the number of Native Americans that a respondent knows is highly correlated to the state in which they reside. On the other hand, doctors are much more likely to know other doctors, regardless of where they live \citep{killworth2006investigating}. The influence that barrier effects can have on model estimates can be reduced by obtaining a representative sample of the population of interest. \cite{killworth2006investigating} cited a study that estimated an unusually high HIV-positive prevalence when compared to national surveys because the original study interviewed only Florida residents, increasing the influence of barrier effects. Unlike transmission error, there does not appear to be any feasible way of estimating barrier effects directly without studying every characteristic of each respondent, since barrier effects depend on both the respondent and the subpopulation considered.



\subsection{Recall Error}
Recall error occurs when respondents inaccurately recall the number of alters they know in a subpopulation \citep{killworth2003two, killworth2006investigating, mccormick2007adjusting, maltiel2015estimating}. In many studies, respondents have only about 30 seconds to recall everyone in their social network that belongs to a subpopulation \citep{killworth2003two}. Thus, respondents will undercount or overestimate the true number of alters. Little if any research has been done one how many people respondents can recall \citep{killworth2003two}. 
Furthermore, given that a respondent has recalled $i$ of $n$ possible alters, the probability that they recall another member decreasing as $i$ increases. Recall bias may also increase as the survey progressing, meaning later questions will suffer from larger recall error \citep{mccarty2001comparing}.

Researchers found that respondents typically overcounted the number of alters in small subpopulations and undercounted the number of alters in large subpopulations \citep{killworth2003two, zheng2006many, mccormick2007adjusting}. 
\cite{killworth2003two} proposed a formula to study the relationship between the estimated degree and the rates at which respondents over and undercount. For large subpopulations, respondents have also been observed to round their responses, typically answering in multiples of 5, but enumerate the individuals in their social network for small subpopulations \citep{mccarty2001comparing, killworth2003two}. For the large subpopulations, respondents relied on ``feel'' \citep{mccarty2001comparing}. This means that even for subpopulations with low transmission error, like those based on names, the recall error may be larger than for other subpopulations. Little work has been done to reduce these recall errors when collecting ARD.

\subsection{Response Bias}
Response bias refers to respondents deliberately misreporting the number of individuals they know in the subpopulation. Respondents may be hesitant to report members of stigmatized subpopulations are in their social network. For example, respondents were hesitant to admit knowing FSWs in household settings \citep{jing2018combining}. Thus, response bias can reasonably be reduced at the data-collection stage by making respondents feel comfortable enough to truthfully answer the survey.

\cite{snidero2009question} studied how the survey design and implementation affected response bias. The authors showed that question order likely did not seem to be an important factor in producing reliable estimates of degree and subpopulation size. However, individual interviewers did have a significant effect on whether interviews were interrupted, refused, or contained missing fields. Therefore, in order to reduce variability of the ARD and provide consistent estimates, it is most important to provide the interviewers with sufficient training.

NSUM was also combined with the randomized response technique (RRT) to reduce response bias (\cite{jing2018combining}). RRT aims to increase the likelihood that a respondent answers sensitive questions by splitting respondents into two groups and randomly asking the respondents either a sensitive or unrelated survey question and ensuring the respondents that only the respondent knows which question they are answering. Since only the respondents knew which question was asked, the researcher must use only the proportion of the sensitive and unrelated questions to calculate the mean response to the sensitive question. 
Respondents were much more likely to answer the sensitive question truthfully under RRT, leading to reliable NSUM estimates. \citep{jing2018combining}. Note, however, that the RRT provided only the average number of FSW that the respondents knew. Thus, the RRT can only be combined with the scale-up estimates that rely on the average number of individuals known in a subpopulation, rather than the number known for each respondent. Based on the current trend of NSU estimators in Section \ref{sec:models}, the RRT is useful only for the most basic estimators.

\section{Models}
\label{sec:models}
In this section, we will discuss the different ARD models. These models can be loosely categorized into three groups: 1) frequentist models which provide subpopulation size estimates, 2) Bayesian models which handle the biases through the prior distributions, and 3) complete network models which focus on estimating network properties using only ARD. Within each subsection, models will be introduced chronologically. Key theoretical properties of the models will also be discussed, although for brevity, full model properties are left to the original publications.


\subsection{Frequentist Models}
\label{sec:basic}

\subsubsection{First NSUM Model}

\cite{johnsen1989estimating} first proposed the Network Scale-up Model to study the number of people who had died in the 1985 Mexico City earthquake. The authors derived bounds for the average network size and point estimates for the unknown subpopulation size and \cite{bernard1991estimating} provided additional empirical results from a larger survey. We focus here on the subpopulation size estimates. While the estimator for subpopulation size is limited in application, it provided a powerful stepping stone for future estimators. The first probability estimator makes no assumption for the distribution of the ARD responses, but notes that the probability of the event that a random respondent knows no one in subpopulation $k$ (denote this event $W_k$) is given by
\begin{equation}
    P(W_k) = \sum_{m = d_{min}}^{d_{max}} P(W_k | y_{ik} = m)P(y_{ik} = m),
\label{eq:pw1}
\end{equation}
where the degrees can vary over the integers from $d_{min}$ to $d_{max}$. The authors then assume that either for a random respondent not in the subpopulation $k$, the respondent's social network is equally likely to have been any subset of size $d_i$ from the general population, or that all subsets of the general population of size $N_k$ were equally likely to be $k$. Using either of these assumptions and several steps of algebra, the authors show that there exists a real number $g$, $1 \leq g \leq d_{max}$, such that
\begin{equation}
    P(W_k) = \sum_{m = d_{min}}^{d_{max}} (1 - N_k / (N - g))^m P(y_{ik} = m) \approx \sum_{m = d_{min}}^{d_{max}} (1 - N_k / N)^m P(y_{ik} = m),
\label{eq:pw2}
\end{equation}
where the approximation holds since $g$ is small with respect to $N$. Now consider (\ref{eq:pw2}) for subpopulations $1$ to $L$, with corresponding values $\epsilon_1 = 1 - N_1 / N$ to $\epsilon_L = 1 - N_L / N$. Since $P(W_k)$ is an increasing function with respect to $\epsilon_k = 1 - N_k/N$, if $1 > \epsilon_1 > \epsilon_2 > \cdots > \epsilon_L$, then $1 > P(W_1) > P(W_2) > \cdots > P(W_L).$ Therefore, if $P(W_u)$ for the unknown subpopulation $u$ is such that $P(W_{j-1}) > P(W_u) > P(W_{j})$, the subpopulation size is also bounded, where
\begin{equation}
    1 - \frac{N_{j-1}}{N} > 1 - \frac{N_k}{N} > 1 - \frac{N_j}{N} \implies \frac{N_{j-1}}{N} < \frac{N_k}{N} < \frac{N_j}{N}.
\end{equation}
Thus, this procedure provides an upper and lower bound for the size of the unknown subpopulation.

\subsubsection{Maximum Likelihood Estimator Models}

In order to derive more precise size estimates, \cite{killworth1998social} proposed a Binomial likelihood based estimator. Overall, six different estimators were proposed to model both personal network size and subpopulation size, but the maximum likelihood based estimator proved the most useful for subpopulation size estimation and was later extended into the most frequently used NSU estimator. The estimator, which we call the \textit{plug-in MLE} (PIMLE), works by first maximizing the following likelihood with respect to $d_i$:
\begin{equation}
    L(d_i; \mathbf{y}) = \prod_{k = 1}^L \binom{d_i}{y_{ik}} \left(\frac{N_k}{N}\right) ^{y_{ik}} \left(1 - \frac{N_k}{N}\right)^{d_i - y_{ik}}.
\label{eq:prob2}
\end{equation}
When $N_k$ are small relative to $N$ and $y_{ik}$ are small relative to $d_i$, the maximum likelihood estimate of $d_i$ is given by
\begin{equation}
    \hat{d}_i = N \cdot \frac{\sum_{k = 1}^L y_{ik}}{\sum_{k = 1}^L N_k}.
\label{eq:kwsocialc}
\end{equation}
Plugging in these $\hat{d}_i$ into (\ref{eq:basic}) yields respondent-level subpopulation estimates $\hat{N}_{u}^{\{i\}} = N \cdot  y_{iu} / \hat{d}_i$. The results are then averaged into a single estimate for $N_{u}$,
\begin{equation}
    \hat{N}_{u}^{PIMLE} = \frac{1}{n}\sum_{i = 1}^n N \cdot  \frac{y_{iu}}{\hat{d}_i},
\label{eq:kwsociale}
\end{equation}
where each term inside the summation is the estimated unknown subpopulation size from respondent $i$. Note that the degree estimate for each respondent depends only on the responses from that respondent and the known subpopulation sizes. Then, each respondent is weighted equally in the final summation. It was shown that $\hat d_i$ is unbiased. Furthermore, Monte Carlo simulations showed $1 / \hat{d}_i$ is essentially unbiased for $1/d_i$ and the back-estimates for $N_{u}$ were essentially unbiased when more than 20 subpopulations with known sizes were used for verification \citep{killworth1998social}. Note that no statement about the bias or standard error of $\hat{N}_{u}^{PIMLE}$ can be made since the final plug-in does not make any distributional assumptions.

\cite{killworth1998estimation} proposed a modified version of the estimate in (\ref{eq:kwsociale}), which has become the most frequently used NSU estimator for unknown subpopulation size. Studies typically refer to this as the \textit{maximum likelihood estimator} (MLE). Instead of back-estimating $N_{u}$ using (\ref{eq:basic}), the MLE method instead maximizes the binomial likelihood
\begin{equation}
        L(N_k; \mathbf{y}, \{d_i\}) = \prod_{i = 1}^n \binom{d_i}{y_{iu}} \left(\frac{N_u}{N}\right) ^{y_{iu}} \left(1 - \frac{N_u}{N}\right)^{d_i - y_{iu}}
\end{equation}
with respect to $N_{u}$, where the $d_i$ are fixed at the estimated $\hat{d_i}$ from (\ref{eq:kwsocialc}). Thus, the final estimate is still uses the known subpopulation to estimate the degrees, but estimates $N_k$ by maximizing a likelihood instead of solving the equality of two ratios. The new MLE for $N_{u}$ is then easily found to be
\begin{equation}
\label{eq:mle}
    \hat{N}_{u}^{MLE} = N \cdot \frac{\sum_{i = 1}^n y_{iu}}{\sum_{i = 1}^n d_i} = \sum_{i = 1}^n y_{iu} \frac{\sum_{k = 1}^L N_k}{\sum_{i = 1}^n \sum_{k = 1}^L y_{ik}}.
\end{equation}
The critical difference in the unknown subpopulation estimate is that the PIMLE averages $N_k$ estimates from each respondent while the MLE maximizes a likelihood using all respondent data simultaneously. The estimate $\hat{N}_{u}$ is unbiased and assuming the $N_u / N$ is sufficiently small, the authors show the standard error is given by
\begin{equation}
    SE(\hat{N}_{u}^{MLE}) = \sqrt{\frac{N \cdot N_{u}}{\sum_{i = 1}^n \hat{d}_i}},
\label{eq:mle_se}
\end{equation}
which decreases as the network sizes $d_i$ increases. When the degrees are small or prevalence is relatively large, the standard error in (\ref{eq:mle_se}) will be inaccurate.

\subsubsection{Weighted Estimators}
\label{sec:habecker}
The MLE estimate of \cite{killworth1998estimation} implicitly values information from large known subpopulations more than small known subpopulations. In order to weight the subpopulations equally, the mean of sums (MoS) estimator first estimates the $d_i$ based on each subpopulation, averages those $\hat{d}_i$'s, and then back-estimates $N_u$ based on the $\hat{d}_i$'s \citep{habecker2015improving}. Thus, the estimate for network size $d_i$ is given by $\hat{d}_i = (N/L) \cdot \sum_{k = 1}^L y_{ik}/N_k$, which then yields the back-estimate for $\hat{N}_u$,
\begin{equation}
\label{eq:hab_N}
    \hat{N}_u^{MoS} = \frac{N}{n} \cdot \sum_{i = 1}^n \frac{y_{iu}}{\hat{d}_i}.
\end{equation}
Note that $\hat{N}_u^{MoS}$ has the same form as $\hat{N}_u^{PIMLE}$, but relies on a different method to calculate the $\hat{d}_i$. The MoS estimator was first proposed in \cite{killworth1998social}, but the authors noted that the variance of the estimate is extremely large when one or more of the known subpopulations are small. \cite{habecker2015improving} proposed controlling this variance by choosing subpopulations of similar size.

\cite{habecker2015improving} also proposed improving the the MLE and the MoS estimator by incorporating weights to adjust for survey characteristics, like probability of selection. Thus, the weighted MLE (\ref{eq:mle}) and the weighted MoS estimator (\ref{eq:hab_N}) are given by
\begin{equation}
    \hat{N}_u^{WMLE} = N \cdot \frac{\sum_{i = 1}^n y_{iu} w_i}{\sum_{i = 1}^n d_i}
\end{equation}
and
\begin{equation}
    \hat{N}_u^{WMoS} = \frac{N}{n} \cdot \sum_{i = 1}^n (y_{iu} w_i) / \hat{d}_i,
\end{equation}
respectively.

\subsubsection{Generalized Scale-up Estimators}
\label{sec:generalized}
\cite{feehan2016generalizing} developed the \textit{generalized scale-up estimator} (GNSUM) to estimate the size of a hard-to-reach subpopulation by using the network property that the total number of in-reports equals the total number of out-reports (i.e. if person $i$ reports they know trait $k$ about person $j$, then person $j$ reports that person $i$ knows trait $k$ about them). The new estimator also requires additional data collected from the hard-to-reach subpopulations called enriched ARD. Enriched ARD is collected by asking members of the unknown population ``How many X's do you know?'' and then ``How many of these X's are aware that you belong to population $G_u$?'' Clearly one limitation of the generalized network scale-up estimator is that it requires directly sampling from the hard-to-reach subpopulation. However, by leveraging the additional information from the enriched ARD, the generalized scale-up estimator can provide more accurate network and subpopulation size estimates.

First, let the hard-to-reach population $G_u$ of size $N_u$ be a subset of the entire population $G$. In practice, researchers sample from $G_F$, which is a separate subset of $G$ called the frame population. Then, let $y_{iu}$ be the number of out-reports from respondent $i$ to subpopulation $G_u$, i.e. how many people respondent $i$ knows in subpopulation $G_u$. Furthermore, let $v_{i,G}$ be the number of in-reports to respondent $i$ from the entire population, also known as the \textit{visibility} of person $i$ to people in $G$. Thus, if we define $y_{G,u} = \sum_{i \in G} y_{iu}$ to be the total number of in-reports and $v_{G,G} = \sum_{i \in G} v_{i, u}$ to be the total number of out-reports, then $y_{G, u} = v_{G,G}$. Multiplying both sides by $N_u$, we can write $N_u = y_{G,u}/(v_{G,G}/N_u).$ This must also be the case for the frame population, i.e. $N_u = y_{F,u}/(v_{G,F}/N_u).$ Then, the estimate $\hat{N}_u^{GNSUM}$ is found by estimating the numerator and denominator separately, i.e.
\begin{equation}
    \hat{N}_u^{GNSUM} = \frac{\hat{y}_{F,u}}{\hat{\bar{v}}_{u, F}}.
\label{eq:feehan3}
\end{equation}
The numerator is found using the ARD, where $\hat{y}_{F,u} = \sum_{i \in s_F} y_{iu}/\pi_i$, $s_F$ is the sample population, and $\pi_i$ is the probability that respondent $i$ is included in the sample from the frame population $G_F$. The enriched ARD is used to find $\hat{\bar{v}}_{u, F}$, but the exact details of the estimator are complicated, and for brevity we leave the description of the estimator to the original manuscript.

\cite{feehan2016generalizing} make several important connections between their model and the MLE estimator. First, the authors show that the GNSUM is equal to the MLE times several adjustment factors, and the MLE is only correct when all adjustments factors are equal to 1 while the generalized scale-up is correct regardless. Procedures for estimating the adjustments factors are provided in the original manuscript. This decomposition also leads an expression for the bias of the MLE. This bias expression can be used to adjust MLE estimates if the adjustment factors are known or estimated.

Recently, \cite{verdery2019estimating} extended the generalized network scale-up estimator to allow venue-based sampling designs instead of the link-tracing samples from the original approach, producing the venue-based generalized scale-up estimator (VB-GNSUM). The generalized NSU estimator is difficult to use with venue-based sampling because estimating $\hat{\bar{v}}_{u, F}$ requires the probability that an individual was included in the sample and because venue-based sampling produces only one sample, rather than the two samples needed for the traditional estimator. \cite{verdery2019estimating} use the same estimator in Equation \ref{eq:feehan3}, but develop new ways to estimate the numerator and denominator.

\subsection{Bayesian Models}
\label{sec:bayesian}

While the original NSUM literature acknowledged that network sizes varied greatly between individuals and the biases influenced the results, the basic models made it difficult to account for these factors and thus relied on averaging estimates. After the initial development of the NSUM, a wave of Bayesian models dominated the topic, allowing the parameters to vary between individuals and subpopulations and relying on posterior estimates. The Bayesian approach inherently allows us to also consider the joint distribution between $N_u$ and the other model parameters.

\subsubsection{The Overdispered Model}
\cite{zheng2006many} proposed the first Bayesian model for ARD, revolutionizing the ARD models. Noting that there was large overdispersion in the data, likely a result from barrier effects and varying network sizes,
\cite{zheng2006many} proposed the overdispersed model, given by
\begin{equation}
    y_{ik} \sim \text{Poisson}(e^{\alpha_i + \beta_k + \gamma_{ik}}).
\end{equation}
The parameter $a_i = e^\alpha_i$ represents the expected degree of respondent $i$, $b_k = e^\beta_k$ represents the proportion of total links that involve subpopulation $k$, and $\gamma_{ik}$ allows for extra variability in the model not accounted for by $\alpha_i$ and $\beta_k$. Thus, if the $\gamma_{ik}$ are constant, the number of people that respondent $i$ knows in subpopulation $k$ is dependent only on the number of people that respondent $i$ knows and the relative prevalence of subpopulation $k$. If, however, $\gamma_{ik}$ varies widely across the $k$, then this suggests the presence of one or more of the ARD biases. The flexibility and hierarchical formulation differentiates the Bayesian models from the frequentist models by allowing for more variation in the responses. It is difficult for the frequentist models to accommodate large changes in how likely each respondent is to know someone in each subpopulation. The addition of an overdispersion parameter allows two respondents with equal degrees to have very different responses, which is often seen in the data.

The authors let $g_{ik} = e^{\gamma_{ik}}$ follow a gamma distribution with mean $1$ and shape parameter $1/(\omega_k - 1)$, which integrated to the following negative binomial model:
\begin{equation}
    y_{ik} \sim \text{negative-binomial}\left(\text{mean} = e^{\alpha_i + \beta_k}, \text{overdispersion} = \omega_{k} \right),
\end{equation}
where $E(y_{ik}) = e^{\alpha_i + \beta_k}$ and $Var(y_{ik}) = \omega_k e^{\alpha_i + \beta_k}$.

Note that the $\alpha_i$'s and $\beta_k$'s are nonidentifiable.
Left untouched, this means that an increase in the expected degree of person $i$ is equivalent to a decrease in the proportion of total links that involve group $k$. The authors chose to leave the nonidentifiability in the model and instead renormalize the $log(\beta_k)$'s using the rare names (those believed to have the least bias) after running the MCMC chain. Full details of the renormalization process can be found in \cite{zheng2006many}. 

The main motivation and utility of this paper is the relationship between the subpopulations and the overdispersion, $\omega_k$. Figure 4 of the original manuscript provides a nice visual summary of the overdispersion estimates. For the considered data-set, the model showed that the ``homeless'' and ``member of Jaycees'' populations had some of the highest overdispersions while the names had some of the lowest. This means that the number of homeless and Jaycees that a respondent knew varied highly between respondents, while the propensity for respondents to know someone named ``Stephanie,'' for example, was roughly equal. The authors pointed out that homeless populations are both ``geographically and socially localized,'' explaining the largest range of propensities between respondents.


\subsubsection{The Latent Profile Models}
\cite{mccormick2010many} note that the normalization procedure of \cite{zheng2006many} does not ensure the degrees are estimated accurately since the transmission errors and the barrier effects can still bias the degree estimates. Thus, the authors propose introducing latent nonrandom mixing to account for the biases. Unlike previous models, the \cite{mccormick2010many} model estimates the propensity for respondents in ego group $e$ to know members of alter group $a$. In their case-study, the authors chose ego and alter groups by crossing gender and age (e.g. males aged 25-64 comprised one ego group), but the ego and alter groups need not match. The initial model is given by
\begin{equation}
\label{eq:mc2010_1}
    y_{ik} \sim \text{negative-binomial}\left(\text{mean} = \mu_{ike}, \text{overdispersion} = \omega'_{k} \right)
\end{equation}
where $\mu_{ike} = d_i \sum_{a = 1}^A m(e,a) N_{ak}/N_a$ is the mean, $N_{ak}/N_a$ is the proportion of subpopulation $k$ within alter group $a$, and $A$ is the total number of alter groups. $m(e,a)$ is the mixing coefficient between group $e$ and alter group $a$, where
\begin{equation}
\label{eq:mc2010_3}
    m(e,a) = \text{E}\left(\frac{d_{ia}}{d_i = \sum_{a = 1}^A d_{ia}} | i \text{ in ego group } e \right),
\end{equation}
and $d_{ia}$ is the number of respondent $i$'s alters that belong to subpopulation $a$. Note that the ego groups and the alter are both exhaustive and mutually exclusive. So for any ego group $e$, we have $\sum_{a=1}^A m(e, a) = 1$. The proposed model in Equations (\ref{eq:mc2010_1})-(\ref{eq:mc2010_3}) accounts for the barrier effects but still suffers from recall bias. To account for recall bias, the authors add a calibration curve derived in \cite{mccormick2007adjusting} to the mean of their negative binomial model, i.e. replace the previous mean with
\begin{equation}
\label{eq:mc2010_4}
    \mu_{ike} = d_i f\left(\sum_{a = 1}^A m(e,a) \frac{N_{ak}}{N_a} \right),
\end{equation}
where the full details of the calibration curve $f(x)$ can be found in the original manuscript.

The authors also provide suggestions for designing future ARD surveys. The \cite{killworth1998social} degree estimates are equivalent to the degree estimates from the above model in expectation if either (1) there is random mixing; or (2) the known subpopulations represent a \textit{scaled-down} population, i.e.
\begin{equation}
    \frac{\sum_{k = 1}^L N_{ak}}{\sum_{k = 1}^K N_k} = \frac{N_a}{N}.
\end{equation}
In words, \cite{mccormick2010many} explain that ``if 20\% of the general population is females under age 30, then 20\% of the people with the [subpopulations] used also must be females under age 30.'' The utility of this requirement is that if the ARD survey is well-designed, then simple models can have the same accuracy as more complicated models. Strategies for designing such a survey and guidelines for understanding the standard error of the estimates are also provided in the original manuscript.

Note, however, that the \cite{mccormick2010many} model does not estimate the size of unknown subpopulations, focusing instead on estimating the individual degree and the distribution of network sizes. In a follow-up paper, \cite{mccormick2012latent} extended the \cite{mccormick2010many} model to estimate the unknown subpopulation sizes via MCMC. Via a two-stage estimation procedure, the authors first use the subpopulations where size is known to fit the model in (\ref{eq:mc2010_1}) and (\ref{eq:mc2010_4}) and then estimate the latent profiles for unknown subpopulation conditional on the estimates values.

\subsubsection{\cite{maltiel2015estimating}}
\label{sec:maltiel}
\cite{maltiel2015estimating} introduced five additional models of increasing complexity to estimate unknown subpopulation sizes from NSUM data and implemented the models in the \texttt{NSUM} \textsf{R} package \citep{maltiel2015NSUM, rlang}. Figure 1 in \cite{maltiel2015estimating} contains a helpful flowchart detailing the four basic proposed models. We discuss only the most complex version of the model and direct the reader to the original manuscript for the other models.

In order to address barrier effects, transmission effects, and recall error, \cite{maltiel2015estimating} proposed the recall adjustment model, given by
\begin{align}
\begin{split}
    y_{ik} &\sim \text{Binom}\left(d_i, e^{r_k} \tau_k q_{ik}\right),
\end{split}
\label{eq:maltiel2}
\end{align}
where $d_i \sim \text{log-normal}(\mu, \sigma^2)$, $r_k \sim N(a + b \log(N_k), \sigma^2_r)$, $q_{ik} \sim \text{Beta}(\text{mean}=m_k, \text{dispersion}=\rho_k)$, and $\tau_k \sim \text{Beta}(\text{mean}=\eta_k, \text{dispersion}=\nu_k)$. The parameters $r_k$ handle the recall error, $q_{ik}$ handle the barrier effects, $\tau_k$ (fixed at 1 for known subpopulations) handle the transmission effects, and $d_i$ represents the varying degree between respondents. Furthermore, the hyperparameters $m_k$ are set to $m_k = N_k / N$. Full prior settings for the hyperparameters are omitted here for clarity. The advantage of this model is that the parameters are clearly separated into the degree and the bias terms. Thus, the response $y_{ik}$ is influenced both by the number of people that respondent $i$ knows as well as the subpopulation size and the biases. Estimating the transmission effect $\tau_k$ for the unknown subpopulation requires some estimate of the visibility factor from additional data sources, for example via the game of contacts.

Ultimately, the authors noted that this complex model was difficult to estimate and proposed removing the $r_k$ parameters and accounting for the recall error post hoc. As others have shown \citep{killworth2003two, zheng2006many, mccormick2007adjusting, mccormick2010many}, respondents seemed to over-report the number of people they knew in small subpopulations and under-report for large subpopulations. The post hoc adjustments are similar to approaches used by others, so details of the adjustments are provided in Section \ref{sec:recall_cal}.

\subsubsection{\cite{teo2019estimating}}
\label{sec:teo}
In order to account for transmission error and barrier effects, \cite{teo2019estimating} proposed two new models that include respondent demographics as regression coefficients. Some of the covariates capture overall respondent characteristics, like age and sex. These covariates can be used to capture additional trends in the response. Other covariates measure the perception that each respondent has for each subpopulation. The main idea is if a respondent views a subpopulation poorly, then they are less likely to know individuals from that subpopulation. The transmission error model assumes a Poisson distribution for the ARD given by
\begin{equation}
    y_{ik} \sim \text{Poisson}\left(\lambda \alpha_i \exp \{\beta_{k} [x_{i,k} - U_{k}] \} N_k \right),
\end{equation}
while transmission error and barrier effect model is then given by
\begin{equation}
    y_{ik} \sim \text{Poisson}\left( \lambda \alpha_i \exp \{\beta_{k} [x_{i,k} - U_{k}] \}\exp \left\{\sum_{j = 1}^p \gamma_{j, k}\mathbf{z}_i \right\}N_k \right),
\end{equation}
where $\mathbf{U}$ represents the upper bound of the Likert scale, $x_{i,k}$ is respondent $i$'s response to the question, and $\mathbf{z} = (\mathbf{z}_1, \ldots, \mathbf{z}_n)$ represents the column-centered covariate matrix. Like \cite{mccormick2010many, mccormick2012latent}, the model relies on respondent characteristics, but treats them more as predictors in a regression framework rather than discretizing the predictors jointly into groups.

\subsection{Complete Network Models}
\label{sec:complete_net}

Full network data is common in a variety of disciplines. For example, economists can use full social networks to determine whether someone is more likely to save more money when another individual monitors their saving progress \citep{breza2019social}. Recently, researchers have focused on using ARD as a substitute for full social network studies since ARD is significantly cheaper and easier to collect. ARD has been estimated to be 70-80\% cheaper to collect that full network surveys \citep{breza2017using}. The question that arises is if and when ARD can be used in place of full social network data. While this line of research deviates significantly from the predominately population size estimation focus of previous methods, we want to make the reader aware of this new research area.

\subsubsection{Latent Surface Model}
\label{sec:surface}
ARD are ultimately partially observed or sampled network data. Thus, \cite{mccormick2015latent} developed a latent surface model, a popular model for complete network data, to analyze incomplete networks like ARD. In this framework, the propensity for person $i$ and $j$ to know each other is proportional to the distance between person $i$ and $j$ in the latent geometry. The latent geometry, in their case, is a $p + 1$-dimensional hypersphere. Furthermore, for subpopulation $G_k$, the collected ARD represents $y_{ik} = \sum_{j \in G_k} \delta_{ij}$, where $\delta_{ij}$ equals 1 if person $i$ and $j$ know each other and 0 otherwise. Thus, if we denote the latent positions of $i$ and $j$ as $z_i$ and $z_{j \in G_k}$, then the distribution of $y_{ik}$ is approximately Poisson distributed with rate $\lambda_{ik} = \sum_{j \in G_k} P(\delta_{ik} = 1 | \bf{Z}_i, \bf{Z_j \in G_k})$. In the complete network case, $j \in G_k$ are observed and known, while for ARD they are unobserved, making it impossible to calculate $\lambda_{ik}$ directly. Instead, the rate is approximated by
\begin{equation}
    \lambda_{ik} \approx N_k \int_{\bf{Z_j \in G_k}} P(\delta_{ik} = 1 | \bf{Z}_i, \bf{Z_j \in G_k}) P(\bf{Z_j \in G_k}) \textit{d} \bf{Z_j \in G_k},
\end{equation}
where $N_k$ is the size of $G_k$. After computing the expectation of the observed data and reparameterizing the model in terms of $d_i$, the likelihood of the latent surface model for ARD is found to be
\begin{equation}
    y_{ik} | d_i, \beta_k, \zeta, \eta_k, \theta_{(z_i, \nu_k)} \sim \text{Poisson}\left(d_i \beta_i \left(\frac{C_{p + 1}(\zeta)C_{p + 1}(\eta_k)}{c_{p + 1}(0)C_{p + 1}\left(\sqrt{\zeta^2 + \eta_k^2 + 2 \zeta \eta_k \cos(\theta_{(z_i, \nu_k)})} \right)} \right) \right),
\end{equation}
where $C_{p + 1}(\cdot)$ is the normalizing constant of the von-Mises Fisher distribution and $\theta_{(z_i, \nu_k)}$ is the angular distance between respondent $i$ and the center of subpopulation $k$. The authors propose a Metropolis MCMC algorithm to sample draws from the posterior distribution.

The latent surface model is closely related to the overdispersed model in \cite{zheng2006many}, and \cite{mccormick2015latent} offer a more detailed comparison of the two models. 
As opposed to the overdispersed model, the latent surface model can put two subpopulations on opposite sides of a latent sphere despite the two subpopulations having similar sizes and dispersion.
\cite{mccormick2015latent} observed from one data set that individuals who reported knowing more people with AIDS also reported knowing more religious individuals. This ability to view the relationship between subpopulations is the main benefit of the latent surface model, but comes at the cost of increased computation.

\subsubsection{Network Statistics}
\label{sec:breza}

\cite{breza2017using} extended the latent surface models and showed that ARD can be used to estimate node- or graph-level statistics under certain situations and provided insight into when ARD is sufficient. Examples of these statistics are individual centrality and average path length of the graph. As the the focus of this paper is remarkably different from our previous discussions, we summarize only their main conclusions and refer the reader to the original manuscript for model details. The authors showed that they could reproduce the findings of complete network studies using only ARD.

\cite{breza2019consistently} further developed the theory behind why and when ARD is sufficient to estimate model parameters for complete networks. One key result from their manuscript is that under certain graphs and given a sufficiently large graph, parameter estimates from ARD are consistent with the true parameters. The authors further develop a system to identify when ARD is sufficient to recover graph statistics. While there are too many results to include here, ARD proves to be an extremely useful tool to estimate graph statistics, especially considering the cost savings and ease of implementation.

\section{Model Calibration}
\label{sec:adjustments}
A significant portion of the NSU literature focuses on calibrating the crude estimates from the NSU models in Section \ref{sec:models} through post hoc adjustments. The calibrations scale the model estimates to correct for ARD biases. Calibrations exist for transmission errors and recall errors, but no adjustments exist for barrier effects because of the aforementioned difficulties in estimating the barrier effects. Note, the original adjustments discussed in this section were typically developed for specific models, but the ideas can be often applied to others, so we discuss them in a general setting.

\subsection{Transmission Calibration}
\label{sec:trans_cal}

The two general approaches to account for transmission error are to use known subpopulations that have low transmission error and to correct crude estimates using the visibility factors. Using only low transmission error subpopulations for estimation does not require any additional data sets and can be applied to most NSU studies. The visibility factor, however, requires an additional sample to estimate and can generally only be used for the hidden subpopulation of interest.

Transmission error can be reduced by using only known subpopulations that are unlikely to have transmission error, like names \citep{mccormick2010many}. It is unlikely that a respondent reports knowing someone without also knowing their name, thereby removing bias from the estimates of degree size. However, the approach does not eliminate the transmission error present in the hard-to-reach subpopulation, where the unknown subpopulation size estimates can still be biased.

More sophisticated methods for selecting the known subpopulations have also been proposed. One approach is to back-estimate each known subpopulation using a leave-one-out procedure and remove all subpopulations which are poorly estimated based on the ratio between the back-estimate and the known size \citep{guo2013estimating}. Similarly, \cite{habecker2015improving} proposed trimming one subpopulation at a time by removing only the worst performing subpopulation ratio in the leave-one-out back-estimates. This step-wise trimming addresses the fact that all model estimates will change after removing any subpopulation, so subpopulations that originally had poor performing ratios might actually perform well after removing one other subpopulation.

The most common method to account for transmission bias is to scale the crude subpopulation size estimates by some scaling factor estimated from an additional data set. The visibility factors are used to directly scale the subpopulation size estimates from the NSU procedure by dividing the crude NSU estimate by the visibility factor. For example, if only 50\% of a female sex worker's social network is aware that they are a female sex worker, than the FSW subpopulation estimate is divided by 0.5 to account for the transmission bias. Given a estimate of the visibility factor, this approach can be easily applied to any estimator. However, only the unknown subpopulation size estimates are scaled and not the degree estimates. Combining the visibility factor method with the only names approach in \cite{mccormick2010many} would likely lead to better degree and subpopulation size estimates.

\subsection{Calibration Curve}
\label{sec:recall_cal}
While \cite{shelley1995knows}, \cite{killworth2003two}, and \cite{shelley2006knows} investigated recall bias in ARD, accounting for the recall bias in the models proved difficult. However, several approaches have been proposed, none of which require any additional data sets. \cite{mccormick2007adjusting} were the first to propose a method to account for recall bias. In order to account for models overestimating small subpopulations and underestimating large subpopulations, the authors constructed a ``calibration curve'' which the authors believed to match the relationship between subpopulation size and recall bias. The calibration curve attempts to scale the \textit{recalled} number of alters to be equal to the true number of alters in subpopulation $k$. As defined in \cite{mccormick2007adjusting}, let $e^{\beta_k}$ be the proportion of ties in the social network for subpopulation $k$, $e^{\beta'_k}$ be for the recalled social network, and then define $\beta'_k = f(\beta_k)$ to be the calibration curve. The calibration curve was first defined as
\begin{equation}
    f(\beta_k) = b + \frac{1}{2}(\beta_k - b) + \frac{1}{2a}\left(1 - e^{-a(\beta_k-b)} \right),
\end{equation}
where $a$ controls how fast the derivative of this curve approaches $1/2$ and $b$ controls at what value of $\beta_k$ the curve changes from correcting for over-reporting to under-reporting. The parameters $a$ and $b$ are then estimated using the subpopulation size estimates without any bias correction.

As mentioned in Section \ref{sec:maltiel}, \cite{maltiel2015estimating} proposed a similar calibration curve. Their calibration works by first treating each known subpopulation as unknown one at a time and estimating the size of that population, $\hat{N}_k$. Then, the errors-in-variables model $\log\left(\hat{N}_k \right) = a + b \log (N_k) + \delta_k + \epsilon_k$ is fit via maximum likelihood to estimate $a$, $b$, and the variances of $\delta_k$ and $\epsilon_k$, where $\delta_k \sim N(0, s_k^2)$ and $\epsilon_k \sim N(0, \sigma^2_\epsilon)$. Estimates of $\log(N_k)$ from the posterior are then transformed using the estimated $a$, $b$, and an additional random noise. The transformation is given by
\begin{equation}
    \frac{Y_k^{[t]}-a}{b} + Z,
\end{equation}
where $Y_k^{[t]}$ is the $t$-th MCMC sample from the posterior distribution of $\log(N_K)$ and $Z \sim N(0, \sigma^2_\epsilon / b^2)$.

\section{Discussion}
\label{sec:discussion}

\begin{table}[!t]
\centering
\caption{Brief summary of NSUM models.}
\label{tab:compare}
\adjustbox{max width=\columnwidth}{
\begin{tabular}{r|rccrc}
\hline \hline
\multicolumn{1}{c}{\textbf{Model}} & \textbf{\begin{tabular}[c]{@{}c@{}}Primary\\[-6pt] Objective\end{tabular}}  & \textbf{\begin{tabular}[c]{@{}c@{}}Requires\\[-6pt] External Data\end{tabular}} &  \textbf{\begin{tabular}[c]{@{}c@{}}Requires Respondent\\[-6pt] Covariates\end{tabular}} & \multicolumn{1}{c}{\textbf{\begin{tabular}[c]{@{}c@{}}Distributional\\[-6pt] Assumptions\end{tabular}}} & \textbf{Bayesian}  \\ \hline
\textbf{\begin{tabular}[l]{@{}l@{}}\cite{johnsen1989estimating} \normalfont{and} \\[-6pt] \cite{bernard1991estimating}\end{tabular}}   &  Degree Estimation    &           &            &    N/A    &     \\ \hline   
\textbf{\begin{tabular}[l]{@{}l@{}}\cite{killworth1998social} \normalfont{and} \\[-6pt] \cite{killworth1998estimation}\end{tabular}}     &   Size Estimation    &         &      &  Binomial   & \\ \hline   
\textbf{\cite{zheng2006many}}      &    Overdispersion/social structure     &                 &                                                                           &  Negative Binomial   &    \\ \hline   
\textbf{\cite{mccormick2010many}}   &   Degree Estimation     &                &       \ding{51}                                                                      &   Negative Binomial    &     \ding{51}   \\\hline   
\textbf{\cite{mccormick2012latent}}     &   Size Estimation     &                 &         \ding{51}                                                                    &    Negative Binomial    &     \ding{51}   \\\hline   
\textbf{\cite{habecker2015improving}}      &    Size Estimation     &                   &                                                                           &    Binomial  &     \\\hline   
\textbf{\cite{maltiel2015estimating}}      &   Size Estimation      &        \ding{51}            &                 &   Binomial    &    \ding{51}  \\\hline   
\textbf{\cite{feehan2016generalizing}}      &   Size Estimation      &         \ding{51}             &   &  N/A    &    \\\hline   
\textbf{\cite{verdery2019estimating}}     &   Size Estimation       &             \ding{51}         &   &  N/A  & \\\hline   
\textbf{\cite{teo2019estimating}}   &     Size Estimation       &                 &    \ding{51}         &    Poisson    &  \ding{51}    \\\hline
\textbf{\cite{mccormick2015latent}}   &    \begin{tabular}[r]{@{}r@{}} Complete \\[-6pt] Network Statistics\end{tabular}       &        &      &   Poisson   &     \ding{51}   \\\hline    
\textbf{\cite{breza2017using}}   &  \begin{tabular}[r]{@{}r@{}} Complete \\[-6pt] Network Statistics\end{tabular}       &                   &        &     Poisson & \ding{51}         \\\hline    
\textbf{\cite{breza2019consistently}}   &     \begin{tabular}[r]{@{}r@{}} Complete \\[-6pt] Network Statistics\end{tabular}       &             &                                                                           &   N/A  &        \\\hline    
\end{tabular}
}
\end{table}

In this manuscript, we discussed the properties of ARD, explored the wide range of models for ARD, and summarized common modifications to these model estimates. For a succinct summary, Table \ref{tab:compare} lists these models, the primary objectives, and additional modeling properties. Many of the models have multiple objectives and estimate several properties, so we record what appears to be the primary objective of the original manuscript. ARD is an increasingly popular survey type for estimating the size of unknown subpopulations due to its relatively cheap method of collecting network data and that individuals from hard-to-reach populations do not need to be surveyed. There are several biases that make modeling ARD difficult, but extensive research has been performed to improve the accuracy and precision of ARD. Recently, ARD has been used in place of full social networks to estimate network properties, and this is a promising area of research that deserved more attention. We include an extensive list of implemented and some proposed NSU studies in the supplementary material. The additional abbreviations we use for the target subpopulations are MCFSW (male client of a female sex worker), MSM (men who have sex with men), RWOS (relationship with opposite sex), EPMS (extra/pre-marital sex), and PED (performance enhancing drugs).

The models for ARD range from very simple MLE estimators to more complex latent surface approaches. Each method has different purposes and assumptions in place, so we refrain from recommending any specific model. Furthermore, the performance of each model has been shown to depend heavily on the sample and place of study. The practical challenge of implementing these methods is also of importance. The frequentist estimators are much easier to implement and have remained the most popular approaches. However, the more complex models offer potential gains in accuracy as well as potentially additional insight into the subpopulations. For example, at the cost of additional computation, the latent surface model proposed in \cite{mccormick2015latent} can place two subpopulations with similar dispersion parameters on opposite sides of the latent surface, indicating that the subpopulations are socially distinct, while the \cite{zheng2006many} model estimates only the dispersion parameters. This is not to say that the latent surface model outperforms the overdispered model, but rather that each model has a different niche. 

ARD, and the extension of NSUM, is a promising area of research, and the models can be extended beyond size estimation. One interesting approach already considered was to invert the problem and estimate the ARD from known subpopulation sizes and degrees. \cite{moody2005fighting} estimated how many people know someone affected by the United States war on terror. Furthermore, there are many hard-to-reach populations that can benefit from the NSUM. \cite{shelton2015proposed} proposed using the NSUM to estimate the prevalence of sex trafficked individuals, noting how difficult it is to obtain a direct estimate. Another promising extension of ARD is to consider other forms of ``How many X's do you know?'' questions. The generalized NSUM estimator requires enriched ARD. Asking additional questions to the respondents may yield helpful information that can be used in a new estimator. Finally, \cite{breza2017using} suggest that ARD should move beyond social networks. For example, the authors consider the question ``How many links does the firm have to firms with trait $k$?'' This question takes a step away from social networks and generalizes ARD to weighted and directed graphs. 

The data and models still have limitations, of course. Estimates for known subpopulation sizes are often significantly incorrect. Approaches for estimating the transmission error are often expensive or impossible for certain subpopulations. Many approaches have reduced response bias, but the success of these methods depends heavily on the unknown subpopulation as well as where the study is implemented. Some papers report eliminating the recall error in the data, but it is difficult to validate these approaches since little research has been done on just recall error. Adjustments are based on leave-one-out estimates of known subpopulations, but the discrepancy between these estimates and the truth are not shown to be due to recall error.

With enough improvement, the NSUM will hopefully drop its ``under-development'' label and be a useful and accurate method for estimating hard-to-reach populations. The models have been used in a large number of real-world studies and have offered promising results in the field of size estimation. The simplicity of the method is attractive and we hope this review inspires new and exciting developments in the field.

\bibliographystyle{apalike}
\bibliography{NSUM_bib}

\begin{thebibliography}{}

\bibitem[Bernard et~al., 2010]{bernard2010counting}
Bernard, H.~R., Hallett, T., Iovita, A., Johnsen, E.~C., Lyerla, R., McCarty,
  C., Mahy, M., Salganik, M.~J., Saliuk, T., Scutelniciuc, O., et~al. (2010).
\newblock Counting hard-to-count populations: the network scale-up method for
  public health.
\newblock {\em Sexually transmitted infections}, 86(Suppl 2):ii11--ii15.

\bibitem[Bernard et~al., 1989]{johnsen1989estimating}
Bernard, H.~R., Johnsen, E.~C., Killworth, P.~D., and Robinson, S. (1989).
\newblock Estimating the size of an average personal network and of an event
  subpopulation.
\newblock In {\em The small world}, pages 159--175. Ablex Press.

\bibitem[Bernard et~al., 1991]{bernard1991estimating}
Bernard, H.~R., Johnsen, E.~C., Killworth, P.~D., and Robinson, S. (1991).
\newblock Estimating the size of an average personal network and of an event
  subpopulation: Some empirical results.
\newblock {\em Social science research}, 20(2):109--121.

\bibitem[BPU and CIF, 2015]{bpu2015population}
BPU, B. and CIF, C. (2015).
\newblock Population size estimation of people who inject drugs in georgia
  2016.
\newblock {\em Tbilisi: CIF}, pages 1--51.

\bibitem[Breza and Chandrasekhar, 2019]{breza2019social}
Breza, E. and Chandrasekhar, A.~G. (2019).
\newblock Social networks, reputation, and commitment: evidence from a savings
  monitors experiment.
\newblock {\em Econometrica}, 87(1):175--216.

\bibitem[Breza et~al., 2017]{breza2017using}
Breza, E., Chandrasekhar, A.~G., McCormick, T.~H., and Pan, M. (2017).
\newblock Using aggregated relational data to feasibly identify network
  structure without network data.
\newblock Technical report, National Bureau of Economic Research.

\bibitem[Breza et~al., 2019]{breza2019consistently}
Breza, E., Chandrasekhar, A.~G., McCormick, T.~H., and Pan, M. (2019).
\newblock Consistently estimating graph statistics using aggregated relational
  data.
\newblock {\em arXiv preprint arXiv:1908.09881}.

\bibitem[Daneshi et~al., 2014]{daneshi2014estimated}
Daneshi, S., Haghdoost, A., Baneshi, M., and Zolala, F. (2014).
\newblock The estimated frequency of spinal cord injury, amputation (hands and
  feet) and death in the bam earthquake using the network scale up method.
\newblock {\em Iranian Journal of Epidemiology}, 10(3):9--14.

\bibitem[Dasgupta et~al., 2014]{dasgupta2014estimating}
Dasgupta, A., Kumar, R., and Sarlos, T. (2014).
\newblock On estimating the average degree.
\newblock In {\em Proceedings of the 23rd international conference on World
  wide web}, pages 795--806.

\bibitem[Ezoe et~al., 2012]{ezoe2012population}
Ezoe, S., Morooka, T., Noda, T., Sabin, M.~L., and Koike, S. (2012).
\newblock Population size estimation of men who have sex with men through the
  network scale-up method in japan.
\newblock {\em PloS one}, 7(1):e31184.

\bibitem[Feehan and Salganik, 2014]{feehan2014networkreporting}
Feehan, D. and Salganik, M. (2014).
\newblock {\em The networkreporting package}.

\bibitem[Feehan and Salganik, 2016]{feehan2016generalizing}
Feehan, D.~M. and Salganik, M.~J. (2016).
\newblock Generalizing the network scale-up method: a new estimator for the
  size of hidden populations.
\newblock {\em Sociological methodology}, 46(1):153--186.

\bibitem[Guo et~al., 2013]{guo2013estimating}
Guo, W., Bao, S., Lin, W., Wu, G., Zhang, W., Hladik, W., Abdul-Quader, A.,
  Bulterys, M., Fuller, S., and Wang, L. (2013).
\newblock Estimating the size of hiv key affected populations in chongqing,
  china, using the network scale-up method.
\newblock {\em PloS one}, 8(8):e71796.

\bibitem[Habecker et~al., 2015]{habecker2015improving}
Habecker, P., Dombrowski, K., and Khan, B. (2015).
\newblock Improving the network scale-up estimator: Incorporating means of
  sums, recursive back estimation, and sampling weights.
\newblock {\em PloS one}, 10(12).

\bibitem[Haghdoost et~al., 2018]{haghdoost2018review}
Haghdoost, A., Gohari, M.~A., Mirzazadeh, A., Zolala, F., and Baneshi, M.~R.
  (2018).
\newblock A review of methods to estimate the visibility factor for bias
  correction in network scale-up studies.
\newblock {\em Epidemiology and health}, 40.

\bibitem[Haghdoost et~al., 2015]{haghdoost2015application}
Haghdoost, A.~A., Baneshi, M.~R., Haji-Maghsoodi, S., Molavi-Vardanjani, H.,
  and Mohebbi, E. (2015).
\newblock Application of a network scale-up method to estimate the size of
  population of breast, ovarian/cervical, prostate and bladder cancers.
\newblock {\em Asian Pac J Cancer Prev}, 16(8):3273--7.

\bibitem[JafariKhounigh et~al., 2014]{jafarikhounigh2014size}
JafariKhounigh, A., Haghdoost, A.~A., SalariLak, S., Zeinalzadeh, A.~H.,
  Yousefi-Farkhad, R., Mohammadzadeh, M., and Holakouie-Naieni, K. (2014).
\newblock Size estimation of most-at-risk groups of hiv/aids using network
  scale-up in tabriz, iran.
\newblock {\em Journal of Clinical Research \& Governance}, 3(1):21--26.

\bibitem[James et~al., 2013]{james2013potential}
James, R.~A., Nepusz, T., Naughton, D.~P., and Petr{\'o}czi, A. (2013).
\newblock A potential inflating effect in estimation models: Cautionary
  evidence from comparing performance enhancing drug and herbal hormonal
  supplement use estimates.
\newblock {\em Psychology of Sport and Exercise}, 14(1):84--96.

\bibitem[Jing et~al., 2018]{jing2018combining}
Jing, L., Lu, Q., Cui, Y., Yu, H., and Wang, T. (2018).
\newblock Combining the randomized response technique and the network scale-up
  method to estimate the female sex worker population size: an exploratory
  study.
\newblock {\em Public health}, 160:81--86.

\bibitem[Jing et~al., 2014]{jing2014estimating}
Jing, L., Qu, C., Yu, H., Wang, T., and Cui, Y. (2014).
\newblock Estimating the sizes of populations at high risk for hiv: a
  comparison study.
\newblock {\em PloS one}, 9(4).

\bibitem[Johnsen et~al., 1995]{johnsen1995social}
Johnsen, E.~C., Bernard, H.~R., Killworth, P.~D., Shelley, G.~A., and McCarty,
  C. (1995).
\newblock A social network approach to corroborating the number of aids/hiv+
  victims in the us.
\newblock {\em Social networks}, 17(3-4):167--187.

\bibitem[Kadushin et~al., 2006]{kadushin2006scale}
Kadushin, C., Killworth, P.~D., Bernard, H.~R., and Beveridge, A.~A. (2006).
\newblock Scale-up methods as applied to estimates of heroin use.
\newblock {\em Journal of Drug Issues}, 36(2):417--440.

\bibitem[Kanato, 2015]{kanato2015size}
Kanato, M. (2015).
\newblock Size estimation of injecting drug users through the network scale-up
  method in thailand.
\newblock {\em Journal of the Medical Association of Thailand= Chotmaihet
  thangphaet}, 98:S17--24.

\bibitem[Kazemzadeh et~al., 2016]{kazemzadeh2016frequency}
Kazemzadeh, Y., Shokoohi, M., Baneshi, M.~R., and Haghdoost, A.~A. (2016).
\newblock The frequency of high-risk behaviors among iranian college students
  using indirect methods: network scale-up and crosswise model.
\newblock {\em International journal of high risk behaviors \& addiction},
  5(3).

\bibitem[Killworth et~al., 1998a]{killworth1998social}
Killworth, P.~D., Johnsen, E.~C., McCarty, C., Shelley, G.~A., and Bernard,
  H.~R. (1998a).
\newblock A social network approach to estimating seroprevalence in the united
  states.
\newblock {\em Social networks}, 20(1):23--50.

\bibitem[Killworth et~al., 2003]{killworth2003two}
Killworth, P.~D., McCarty, C., Bernard, H.~R., Johnsen, E.~C., Domini, J., and
  Shelley, G.~A. (2003).
\newblock Two interpretations of reports of knowledge of subpopulation sizes.
\newblock {\em Social networks}, 25(2):141--160.

\bibitem[Killworth et~al., 1998b]{killworth1998estimation}
Killworth, P.~D., McCarty, C., Bernard, H.~R., Shelley, G.~A., and Johnsen,
  E.~C. (1998b).
\newblock Estimation of seroprevalence, rape, and homelessness in the united
  states using a social network approach.
\newblock {\em Evaluation review}, 22(2):289--308.

\bibitem[Killworth et~al., 2006]{killworth2006investigating}
Killworth, P.~D., McCarty, C., Johnsen, E.~C., Bernard, H.~R., and Shelley,
  G.~A. (2006).
\newblock Investigating the variation of personal network size under unknown
  error conditions.
\newblock {\em Sociological Methods \& Research}, 35(1):84--112.

\bibitem[Maghsoudi et~al., 2014]{maghsoudi2014network}
Maghsoudi, A., Baneshi, M.~R., Neydavoodi, M., and Haghdoost, A. (2014).
\newblock Network scale-up correction factors for population size estimation of
  people who inject drugs and female sex workers in iran.
\newblock {\em PloS one}, 9(11):e110917.

\bibitem[Maltiel and Baraff, 2015]{maltiel2015NSUM}
Maltiel, R. and Baraff, A.~J. (2015).
\newblock {\em NSUM: Network Scale Up Method}.
\newblock R package version 1.0.

\bibitem[Maltiel et~al., 2015]{maltiel2015estimating}
Maltiel, R., Raftery, A.~E., McCormick, T.~H., and Baraff, A.~J. (2015).
\newblock Estimating population size using the network scale up method.
\newblock {\em The annals of applied statistics}, 9(3):1247.

\bibitem[McCarty et~al., 2001]{mccarty2001comparing}
McCarty, C., Killworth, P.~D., Bernard, H.~R., Johnsen, E.~C., and Shelley,
  G.~A. (2001).
\newblock Comparing two methods for estimating network size.
\newblock {\em Human organization}, 60(1):28--39.

\bibitem[McCormick et~al., 2010]{mccormick2010many}
McCormick, T.~H., Salganik, M.~J., and Zheng, T. (2010).
\newblock How many people do you know?: Efficiently estimating personal network
  size.
\newblock {\em Journal of the American Statistical Association},
  105(489):59--70.

\bibitem[McCormick and Zheng, 2012]{mccormick2012latent}
McCormick, T.~H. and Zheng, T. (2012).
\newblock Latent demographic profile estimation in hard-to-reach groups.
\newblock {\em The annals of applied statistics}, 6(4):1795.

\bibitem[McCormick and Zheng, 2015]{mccormick2015latent}
McCormick, T.~H. and Zheng, T. (2015).
\newblock Latent surface models for networks using aggregated relational data.
\newblock {\em Journal of the American Statistical Association},
  110(512):1684--1695.

\bibitem[McCormick et~al., 2007]{mccormick2007adjusting}
McCormick, T.~H., Zheng, T., et~al. (2007).
\newblock Adjusting for recall bias in “how many x’s do you know?”
  surveys.
\newblock In {\em Proceedings of the joint statistical meetings}.

\bibitem[Mohebbi et~al., 2014]{mohebbi2014application}
Mohebbi, E., Baneshi, M.~R., Haji-Maghsoodi, S., and Haghdoost, A.~A. (2014).
\newblock The application of network scale up method on estimating the
  prevalence of some disabilities in the southeast of iran.
\newblock {\em Journal of research in health sciences}, 14(4):272--275.

\bibitem[Moody, 2005]{moody2005fighting}
Moody, J. (2005).
\newblock Fighting a hydra: A note on the network embeddedness of the war on
  terror.
\newblock {\em Structure and Dynamics}, 1(2).

\bibitem[Narouee et~al., 2019]{narouee2019size}
Narouee, S., Shati, M., Nasehi, M., and Dadgar, F. (2019).
\newblock The size estimation of injection drug users (idus) using the network
  scale-up method (nsum) in iranshahr, iran.
\newblock {\em Medical Journal of The Islamic Republic of Iran (MJIRI)},
  33(1):972--976.

\bibitem[Nasiri et~al., 2019]{nasiri2019population}
Nasiri, N., Abedi, L., Hajebi, A., Noroozi, A., Khalili, M., Chegeni, M., Nili,
  S., Taheri-Soodejani, M., Noroozi, M., Shahesmaeili, A., et~al. (2019).
\newblock Population size estimation of tramadol misusers in urban population
  in iran: Synthesis of methods and results.
\newblock {\em Addiction \& Health}, 11(3):173.

\bibitem[Nikfarjam et~al., 2017]{nikfarjam2017frequency}
Nikfarjam, A., Hajimaghsoudi, S., Rastegari, A., Haghdoost, A.~A., Nasehi,
  A.~A., Memaryan, N., Tarjoman, T., and Baneshi, M.~R. (2017).
\newblock The frequency of alcohol use in iranian urban population: the results
  of a national network scale up survey.
\newblock {\em International journal of health policy and management}, 6(2):97.

\bibitem[Nikfarjam et~al., 2016]{nikfarjam2016national}
Nikfarjam, A., Shokoohi, M., Shahesmaeili, A., Haghdoost, A.~A., Baneshi,
  M.~R., Haji-Maghsoudi, S., Rastegari, A., Nasehi, A.~A., Memaryan, N., and
  Tarjoman, T. (2016).
\newblock National population size estimation of illicit drug users through the
  network scale-up method in 2013 in iran.
\newblock {\em International Journal of Drug Policy}, 31:147--152.

\bibitem[{R Core Team}, 2019]{rlang}
{R Core Team} (2019).
\newblock {\em R: A Language and Environment for Statistical Computing}.
\newblock R Foundation for Statistical Computing, Vienna, Austria.

\bibitem[Rastegari et~al., 2014]{rastegari2014estimating}
Rastegari, A., Baneshi, M.~R., Haji-Maghsoudi, S., Nakhaee, N., Eslami, M.,
  Malekafzali, H., and Haghdoost, A.~A. (2014).
\newblock Estimating the annual incidence of abortions in iran applying a
  network scale-up approach.
\newblock {\em Iranian Red Crescent Medical Journal}, 16(10).

\bibitem[Rastegari et~al., 2013]{rastegari2013estimation}
Rastegari, A., Haji-Maghsoudi, S., Haghdoost, A., Shatti, M., Tarjoman, T., and
  Baneshi, M.~R. (2013).
\newblock The estimation of active social network size of the iranian
  population.
\newblock {\em Global journal of health science}, 5(4):217.

\bibitem[Rogerson, 1997]{rogerson1997estimating}
Rogerson, P.~A. (1997).
\newblock Estimating the size of social networks.
\newblock {\em Geographical Analysis}, 29(1):50--63.

\bibitem[{Rwanda Biomedical Center}, 2012]{center2012estimating}
{Rwanda Biomedical Center} (2012).
\newblock Estimating the size of key populations at higher risk of hiv through
  a household survey (esphs) rwanda 2011.
\newblock Technical report, Technical Report. Calverton, MD: RBC/IHDPC, SPF,
  UNAIDS, and ICF International.

\bibitem[Sabin et~al., 2016]{sabin2016availability}
Sabin, K., Zhao, J., Calleja, J. M.~G., Sheng, Y., Garcia, S.~A., Reinisch, A.,
  and Komatsu, R. (2016).
\newblock Availability and quality of size estimations of female sex workers,
  men who have sex with men, people who inject drugs and transgender women in
  low-and middle-income countries.
\newblock {\em PLoS One}, 11(5).

\bibitem[Sajjadi et~al., 2018]{sajjadi2018indirect}
Sajjadi, H., Shushtari, Z.~J., Shati, M., Salimi, Y., Dejman, M., Vameghi, M.,
  Karimi, S., and Mahmoodi, Z. (2018).
\newblock An indirect estimation of the population size of students with
  high-risk behaviors in select universities of medical sciences: A network
  scale-up study.
\newblock {\em PloS one}, 13(5).

\bibitem[Salganik et~al., 2011a]{salganik2011assessing}
Salganik, M.~J., Fazito, D., Bertoni, N., Abdo, A.~H., Mello, M.~B., and
  Bastos, F.~I. (2011a).
\newblock Assessing network scale-up estimates for groups most at risk of
  hiv/aids: evidence from a multiple-method study of heavy drug users in
  curitiba, brazil.
\newblock {\em American journal of epidemiology}, 174(10):1190--1196.

\bibitem[Salganik et~al., 2011b]{salganik2011game}
Salganik, M.~J., Mello, M.~B., Abdo, A.~H., Bertoni, N., Fazito, D., and
  Bastos, F.~I. (2011b).
\newblock The game of contacts: estimating the social visibility of groups.
\newblock {\em Social networks}, 33(1):70--78.

\bibitem[Sharifi et~al., 2017]{sharifi2017population}
Sharifi, H., Karamouzian, M., Baneshi, M.~R., Shokoohi, M., Haghdoost, A.,
  McFarland, W., and Mirzazadeh, A. (2017).
\newblock Population size estimation of female sex workers in iran: Synthesis
  of methods and results.
\newblock {\em PloS one}, 12(8).

\bibitem[Shati et~al., 2014]{shati2014social}
Shati, M., Haghdoost, A., Majdzadeh, R., Mohammad, K., and Mortazavi, S.
  (2014).
\newblock Social network size estimation and determinants in tehran province
  residents.
\newblock {\em Iranian journal of public health}, 43(8):1079.

\bibitem[Sheikhzadeh et~al., 2016]{sheikhzadeh2016comparing}
Sheikhzadeh, K., Baneshi, M.~R., Afshari, M., and Haghdoost, A.~A. (2016).
\newblock Comparing direct, network scale-up, and proxy respondent methods in
  estimating risky behaviors among collegians.
\newblock {\em Journal of Substance Use}, 21(1):9--13.

\bibitem[Shelley et~al., 1995]{shelley1995knows}
Shelley, G.~A., Bernard, H.~R., Killworth, P., Johnsen, E., and McCarty, C.
  (1995).
\newblock Who knows your hiv status? what hiv+ patients and their network
  members know about each other.
\newblock {\em Social networks}, 17(3-4):189--217.

\bibitem[Shelley et~al., 2006]{shelley2006knows}
Shelley, G.~A., Killworth, P.~D., Bernard, H.~R., McCarty, C., Johnsen, E.~C.,
  and Rice, R.~E. (2006).
\newblock Who knows your hiv status ii?: Information propagation within social
  networks of seropositive people.
\newblock {\em Human Organization}, 65(4):430.

\bibitem[Shelton, 2015]{shelton2015proposed}
Shelton, J.~F. (2015).
\newblock Proposed utilization of the network scale-up method to estimate the
  prevalence of trafficked persons.
\newblock In {\em Forum on Crime \& Society}, volume~8.

\bibitem[Shokoohi et~al., 2010]{shokoohi2010estimation}
Shokoohi, M., Baneshi, M.~R., and Haghdoost, A.~A. (2010).
\newblock Estimation of the active network size of kermanian males.
\newblock {\em Addiction \& health}, 2(3-4):81.

\bibitem[Shokoohi et~al., 2012]{shokoohi2012size}
Shokoohi, M., Baneshi, M.~R., and Haghdoost, A.-A. (2012).
\newblock Size estimation of groups at high risk of hiv/aids using network
  scale up in kerman, iran.
\newblock {\em International Journal of Preventive Medicine}, 3(7):471.

\bibitem[Snidero et~al., 2007]{snidero2007use}
Snidero, S., Morra, B., Corradetti, R., and Gregori, D. (2007).
\newblock Use of the scale-up methods in injury prevention research: An
  empirical assessment to the case of choking in children.
\newblock {\em Social Networks}, 29(4):527--538.

\bibitem[Snidero et~al., 2012]{snidero2012scale}
Snidero, S., Soriani, N., Baldi, I., Zobec, F., Berchialla, P., and Gregori, D.
  (2012).
\newblock Scale-up approach in cati surveys for estimating the number of
  foreign body injuries in the aero-digestive tract in children.
\newblock {\em International journal of environmental research and public
  health}, 9(11):4056--4067.

\bibitem[Snidero et~al., 2009]{snidero2009question}
Snidero, S., Zobec, F., Berchialla, P., Corradetti, R., and Gregori, D. (2009).
\newblock Question order and interviewer effects in cati scale-up surveys.
\newblock {\em Sociological Methods \& Research}, 38(2):287--305.

\bibitem[Sulaberidze et~al., 2016]{sulaberidze2016population}
Sulaberidze, L., Mirzazadeh, A., Chikovani, I., Shengelia, N., Tsereteli, N.,
  and Gotsadze, G. (2016).
\newblock Population size estimation of men who have sex with men in tbilisi,
  georgia; multiple methods and triangulation of findings.
\newblock {\em PloS one}, 11(2).

\bibitem[Teo et~al., 2019]{teo2019estimating}
Teo, A. K.~J., Prem, K., Chen, M.~I., Roellin, A., Wong, M.~L., La, H.~H., and
  Cook, A.~R. (2019).
\newblock Estimating the size of key populations for hiv in singapore using the
  network scale-up method.
\newblock {\em Sexually transmitted infections}, 95(8):602--607.

\bibitem[{UNAIDS and WHO}, 2010]{unaids2010guidelines}
{UNAIDS and WHO} (2010).
\newblock Guidelines on estimating the size of populations most at risk to hiv.
\newblock {\em Geneva, Switzerland: World Health Organization}, page~51.

\bibitem[Vardanjani et~al., 2015]{vardanjani2015cancer}
Vardanjani, H.~M., Baneshi, M.~R., and Haghdoost, A. (2015).
\newblock Cancer visibility among iranian familial networks: to what extent can
  we rely on family history reports?
\newblock {\em PloS one}, 10(8).

\bibitem[Verdery et~al., 2019]{verdery2019estimating}
Verdery, A.~M., Weir, S., Reynolds, Z., Mulholland, G., and Edwards, J.~K.
  (2019).
\newblock Estimating hidden population sizes with venue-based sampling:
  Extensions of the generalized network scale-up estimator.
\newblock {\em Epidemiology}, 30(6):901--910.

\bibitem[Wang et~al., 2015]{wang2015application}
Wang, J., Yang, Y., Zhao, W., Su, H., Zhao, Y., Chen, Y., Zhang, T., and Zhang,
  T. (2015).
\newblock Application of network scale up method in the estimation of
  population size for men who have sex with men in shanghai, china.
\newblock {\em PloS one}, 10(11).

\bibitem[Yang and Yang, 2017]{yang2017estimating}
Yang, X.~Y. and Yang, F. (2017).
\newblock Estimating religious populations with the network scale-up method: A
  practical alternative to self-report.
\newblock {\em Journal for the Scientific Study of Religion}, 56(4):703--719.

\bibitem[Zahedi et~al., 2018]{zahedi2018self}
Zahedi, R., Noroozi, A., Hajebi, A., Haghdoost, A.~A., Baneshi, M.~R., Sharifi,
  H., and Mirzazadeh, A. (2018).
\newblock Self-reported and network scale-up estimates of substance use
  prevalence among university students in kerman, iran.
\newblock {\em Journal of research in health sciences}, 18(2).

\bibitem[Zamanian et~al., 2016]{zamanian2016estimating}
Zamanian, M., Baneshi, M.~R., Haghdoost, A., and Zolala, F. (2016).
\newblock Estimating the visibility rate of abortion: a case study of kerman,
  iran.
\newblock {\em BMJ open}, 6(10):e012761.

\bibitem[Zamanian et~al., 2019]{zamanian2019methodological}
Zamanian, M., Zolala, F., Haghdoost, A.~A., Haji-Maghsoudi, S., Heydari, Z.,
  and Baneshi, M.~R. (2019).
\newblock Methodological considerations in using the network scale up (nsu) for
  the estimation of risky behaviors of particular age-gender groups: An example
  in the case of intentional abortion.
\newblock {\em PloS one}, 14(6).

\bibitem[Zheng et~al., 2006]{zheng2006many}
Zheng, T., Salganik, M.~J., and Gelman, A. (2006).
\newblock How many people do you know in prison? using overdispersion in count
  data to estimate social structure in networks.
\newblock {\em Journal of the American Statistical Association},
  101(474):409--423.

\end{thebibliography}

\appendix
\appendixpage
\section{Table of Network Scale-up Method Studies}

\begin{longtable}{lll}
\caption{A survey of NSUM applied studies. Country and year of study are recorded when provided. When the year was not provided, the publication year is denoted by *.}\\
\label{tab:NSU_survey}
Subpopulation & Country, year &  References \\ \hline
\endfirsthead
\caption{(continued)} \\ \hline
\endhead
    FSW &  Rwanda, 2011     &   \cite{center2012estimating}  \\
        &   China, 2011   & \cite{guo2013estimating} \\
        &  China, 2012     &  \cite{jing2018combining}   \\
        &  Iran, 2014*  &  \cite{maghsoudi2014network} \\
        &  Iran, 2014     &  \cite{sharifi2017population}   \\
        &  Iran, 2014*     &   \cite{jafarikhounigh2014size}  \\
        & Singapore, 2017 & \cite{teo2019estimating} \\
    Drug Users    &   United States, 1997    &   \cite{kadushin2006scale}  \\
        &   Brazil, 2009    & \cite{salganik2011game}    \\
        &   Brazil, 2009-2010    &  \cite{salganik2011assessing}   \\
        &   Rwanda, 2011    &  \cite{center2012estimating}   \\
        &   China, 2011   & \cite{guo2013estimating} \\
        &   Iran, 2012*    &   \cite{shokoohi2012size}  \\
        &   Iran, 2014*    &   \cite{maghsoudi2014network}  \\
        &   Iran, 2016    &  \cite{nasiri2019population}   \\
        &   Iran, 2013    &  \cite{nikfarjam2016national}   \\
        &   Iran, 2012-2013    & \cite{kazemzadeh2016frequency}    \\
        &   Iran, 2016*    &  \cite{sheikhzadeh2016comparing}   \\
        &   Iran, 2016    &   \cite{zahedi2018self}  \\
        &   Iran, 2014*    &  \cite{jafarikhounigh2014size}   \\
        &   Iran, 2015    &  \cite{sajjadi2018indirect}   \\
        &   Iran, 2016-2017   &   \cite{narouee2019size}  \\
        &   Thailand, 2014    &  \cite{kanato2015size}   \\
        &   Georgia, 2014   &  \cite{bpu2015population}   \\
        &   Singapore, 2017    &  \cite{teo2019estimating}   \\
    MCFSW   &   Rwanda, 2011    &   \cite{center2012estimating}  \\
        &   China, 2011   & \cite{guo2013estimating} \\
        &   Iran, 2014*    &   \cite{jafarikhounigh2014size}  \\
        &   Iran, 2015    &   \cite{sajjadi2018indirect}  \\
        &   Singapore, 2017    &  \cite{teo2019estimating}   \\
    MSM    &   Japan, 2009    &  \cite{ezoe2012population}   \\
        &   Rwanda, 2011   &  \cite{center2012estimating}   \\
        &   China, 2011   & \cite{guo2013estimating} \\
        &   China, 2012    &   \cite{jing2014estimating}  \\
        &   China, 2012    &   \cite{wang2015application}  \\
        &   Georgia, 2014    &   \cite{sulaberidze2016population}  \\
        &   Iran, 2014*    &   \cite{jafarikhounigh2014size}  \\
        &   Iran, 2015    &   \cite{sajjadi2018indirect}  \\
        &   Singapore, 2017    &   \cite{teo2019estimating}  \\
    Abortion    &    Iran, 2012    &   \cite{rastegari2014estimating}  \\
        &   Iran, 2015    &   \cite{zamanian2016estimating}  \\
        &   Iran, 2016    &   \cite{zamanian2019methodological}  \\
    Alcohol    &   Iran, 2012    &   \cite{nikfarjam2017frequency}  \\
        &   Iran, 2012-2013    &   \cite{kazemzadeh2016frequency}  \\
        &  Iran, 2014*     &   \cite{jafarikhounigh2014size}  \\
        &   Iran, 2015    &   \cite{sajjadi2018indirect}  \\
        &   Iran, 2016*    &  \cite{sheikhzadeh2016comparing}   \\
    Cancer    &   Iran, 2012-2013    &   \cite{haghdoost2015application}  \\
        &   Iran, 2014    &   \cite{vardanjani2015cancer}  \\
    Choking    &    Italy, 2004   &   \cite{snidero2007use}  \\
        &   Italy, 2004    &   \cite{snidero2012scale}  \\
    Social Network Size    &   Iran, 2010*    &   \cite{shokoohi2010estimation}  \\
        &   Iran, 2012    &   \cite{shati2014social}  \\
        &   Iran, 2013*     &   \cite{rastegari2013estimation}  \\
    Disabilities    &   Iran, 2012     &   \cite{mohebbi2014application}  \\
    Religious Groups    &   United States, 2016    &   \cite{yang2017estimating}  \\
    Seroprevalence    &    United States, 1994    &   \cite{killworth1998estimation}  \\
    Rape    &    United States, 1994   &   \cite{killworth1998estimation}  \\
    Homelessness    &    United States, 1994   &   \cite{killworth1998estimation}  \\
    RWOS    &   Iran, 2012-2013    &   \cite{kazemzadeh2016frequency}  \\
    EPMS     &   Iran, 2012-2013    &   \cite{kazemzadeh2016frequency}  \\
        &    Iran, 2014*   &   \cite{jafarikhounigh2014size}  \\
        &    Iran, 2015   &   \cite{sajjadi2018indirect}  \\
        &   Iran, 2016*    &  \cite{sheikhzadeh2016comparing}   \\
    PED    &   UK and Southern Ireland, 2013*    &   \cite{james2013potential}  \\
    Sex Trafficked    &   United States    &   \cite{shelton2015proposed}  \\
    Suicide    &    Iran, 2014   &   \cite{daneshi2014estimated}  \\
\end{longtable}

\end{document}